\documentclass[letter]{aa} 
\usepackage{graphicx}
\usepackage{txfonts}
\usepackage{natbib,9200defs}
\begin{document}
   \title{Black hole growth and stellar assembly at high-z}


   \author{M. Polletta
          \inst{1,2}
          }

   \offprints{M. Polletta}

   \institute{$^1$Institut d'Astrophysique de Paris (IAP), 
              98bis boulevard Arago,
              Paris, 75014 France\\
              $^2$INAF - IASF Milano, 
              via E. Bassini 15,
              Milan, 20133 Italy\\
              \email{polletta@iap.fr}
             }

   \date{Received 6 December 2007/ Accepted 4 February 2008}

   \abstract
   {Observations indicate a strong link between star formation and
    black hole (BH) growth, but some questions remain unanswered: whether
    both activities are coeval or whether one precedes the other, what their
    characteristic timescales are, and what kinds of physical processes are
    responsible for this interplay.}
   {We examine stellar and BH masses (M$_{*}$ and M$_{BH}$) in $z$$\sim$2
    active systems at the peak of their AGN or star formation activity to
    investigate how they are linked and whether AGN radiative or else radio
    power provides a feedback mechanism that regulates the stellar growth in
    these systems.}
   {We analyze the infrared (IR) spectral energy distributions
    of radio, sub-millimeter and mid-IR selected AGNs at
    $z\sim$1--3 and constrain their stellar and AGN luminosities 
    using AGN and host-galaxy templates.}
   {We find evidence of increasing stellar light, thereby decreasing the AGN
    mid-IR power going from mid-IR selected AGNs, to radio galaxies, and to
    sub-millimeter AGNs. This trend can be explained by either decreasing
    Eddington ratios or increasing offsets from the local M$_{BH}$--M$_{*}$
    relation. All systems are characterized by high star formation rates
    regardless of their different AGN powers, thus neither AGN radiative power
    nor AGN-driven radio activity seems to influence the star formation rate
    in the selected AGNs. We discuss two possible evolutionary scenarios
    that might link these three AGN classes.}
   {}


   \maketitle
%

\section{Introduction}

A new paradigm has recently emerged in which galaxies and their black holes
(BHs) grow in tandem, and BHs are responsible for regulating star formation
(SF) through feedback processes and thus shaping many of the basic
relationships between galaxy
properties~\citep[e.g.][]{croton06b,merloni07,balogh01,lin03,somerville04}.
Feedback processes are thought to be driven by AGNs, and they manifest
themselves in outflowing winds or radio jets that can heat a significant
fraction of the gas even, from the most massive dark-matter halos, and
terminate the SF process in galaxies~\citep{dimatteo05,croton06b}.

Current evolutionary models predict that AGN-feedback takes place in obscured
and extremely luminous AGNs~\citep{hopkins05a} or in radio-loud
AGNs~\citep{croton06a}, especially at $z\gtrsim$2 where both BH
accretion and SF activity
peak~\citep[e.g.][]{giavalisco04,marconi04}.  Although not in large numbers,
these kinds of objects have been found in wide-area radio and infrared
surveys, where the effects of obscuration are minimized and AGN signatures
are clearly identified. Moreover, thanks to the advent of the {\it Spitzer
Space Telescope}, it has become possible to study the multi-wavelength
properties of these objects in
detail~\citep{ogle06,seymour07,sajina07b,polletta08}.

In this Letter, we compare the relative luminosity produced by stars and by
AGN-heated dust in three classes of AGNs, high-$z$ radio galaxies
(HzRGs hereinafter), sub-millimeter-detected AGNs (SMG/AGNs hereinafter),
and IR-selected obscured QSOs (IsOQs hereinafter), to investigate whether
these three classes represent members of the same population caught during
different evolutionary stages, or whether they are the result of different
histories and physical conditions.

\section{Obscured AGNs at $z\sim$2}

HzRGs have been discovered due to to their large radio fluxes
($\gtrsim$10\,mJy) in shallow wide surveys~\citep[e.g.  NVSS,
WENSS][]{condon98,rengelink97}. Since they are extremely rare, it is
necessary to sample large volumes to find them. They are usually
characterized by high excitation emission lines in the optical and strong
mid-IR (MIR) emission, and their radio emission is dominated by non-thermal
radiation from radio lobes. In the X-ray, they show both the contribution
from a jet and an often absorbed component associated with the accretion
disk~\citep[e.g.][]{seymour07,hardcastle06}.

IsOQs have been been discovered as bright ($\gtrsim$1\,mJy) 24$\mu$m sources
in \spitzer\ surveys with faint optical counterparts~\citep[$F_{24\mu
m}/F_R$$>$200;][]{houck04,yan05}. Their MIR SEDs are usually consistent
with power-law or convex spectral shapes rapidly rising towards longer
wavelengths, and they often display absorption due to
silicates~\citep{alonso06,weedman06a,polletta06,polletta08}.

SMG/AGNs were first discovered in sub-millimeter surveys, and
their AGN nature was revealed by their X-ray emission~\citep{alexander05a}.
X-ray observations indicate that these galaxies host moderately luminous
and heavily absorbed AGNs. Their optical-IR emission is mainly dominated by
the host-galaxy, and they are characterized by intense starburst
activity~\citep{borys05,alexander05a,alexander05b}.

For this work we selected 17 HzRGs from~\citet{seymour07}, 11 X-ray detected
SMG/AGNs from~\citet{borys05}, and 21 IsOQs from~\citet{polletta08}. The
three samples only include sources at 1$<$$z$$<$3. The median and redshift
range of each sample are reported in Table~\ref{radio_tab}. All luminosities
were derived assuming H$_0$ = 70\kmsMpc, $\Omega_{M}$=0.3, and
$\Omega_{\Lambda}$=0.7. Throughout the paper, the uncertainties associated
to the median values correspond to the average deviation from the median.

\subsection{Host stellar light and AGN accretion power}

\begin{figure}
   \centering
   \includegraphics[width=9cm]{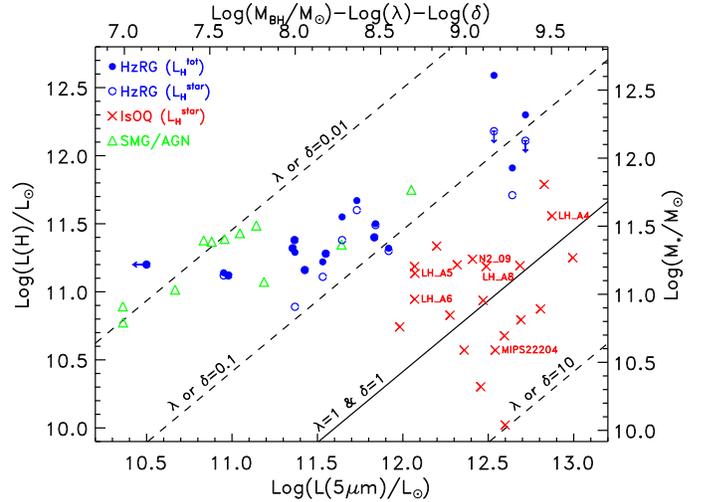}
      \caption{Comparison between the H-band (1.6\,$\mu$m) and the MIR
(5\,$\mu$m) luminosities of three classes of AGNs,
HzRGs~\citep{seymour07} ({\it full and open blue circles}),
SMG/AGNs~\citep{borys05} ({\it green triangles}), and IsOQs~\citep{polletta08}
({\it red crosses}). The H-band luminosity of the HzRGs refers to the total
({\it full blue circles}) or stellar-only ({\it open blue circles}) luminosity. Downward
pointing arrows represent upper limits to the H-band luminosity, and left
pointing arrows indicate upper limits to the MIR luminosity.  The diagonal lines
represent the expected luminosities for BHs of various masses, as indicated
on the top axis, assuming L$^{AGN}_{bol}$/L$_{Edd}$=$\lambda$, where 
L$^{AGN}_{bol}$ is derived from L(5\,\micron)~\citep{polletta08}, and the local
M$_{BH}$-M$_{*}$ relation~\citep{marconi03} where M$_{*}$ is derived from
L(H)~\citep{seymour07} (see right vertical axis) with an offset $\delta$ so
that higher or lower BH masses are derived for, respectively, $\delta$$>$1
or $<$1. The names of radio-detected IsOQs are annotated just to the right
of their location (see Table~\ref{radio_tab}).}
         \label{lh_lmir}
   \end{figure}

A common problem in AGN studies is to estimate the host-galaxy contribution
to the bolometric luminosity and, especially, the stellar mass.  Since the
stellar emission peaks in the near-IR (NIR), typically at 1.6\,\micron\ in
the rest frame (or H-band), and the ratio between NIR luminosity and stellar
mass (M$_{*}$) are characterized by little dispersion, the NIR luminosity is
a proxy of the bulge mass or the luminosity of the host-galaxy. Note that
the L(NIR)/M$_{*}$ depends on the SF history and age of the system, e.g.
younger stellar populations can yield 30\% lower stellar mass estimates than
older stellar populations~\citep[e.g.][]{seymour07}. Since all selected
objects are at similar redshifts and their AGN component does not
dominate at NIR wavelengths, reasonable estimates of stellar emission and
mass can be obtained from L(NIR). In IsOQs, the NIR stellar light is
estimated after subtracting the AGN component. We used the NIR estimates
from~\citet{polletta08}, where the IR SEDs are modeled with a combination of
host and torus models. For the HzRGs, we used the estimates
from~\citet{seymour07} where the contribution from AGN-heated dust to the
NIR is also removed. NIR luminosities were derived for the SMG/AGNs by
fitting their SED~\citep{borys05} with galaxy templates. Stellar masses were
derived from L(H) following the method in~\citet{seymour07}. The estimated
NIR luminosities and corresponding stellar masses are reported in
Figure~\ref{lh_lmir}.

The IR SEDs of HzRGs and IsOQs indicate that their MIR emission is dominated
by AGN-heated dust. Since the MIR (i.e., at 5\,$\mu$m in the rest frame) is
not significantly affected by dust extinction, and this is where the 
AGN-heated dust radiation peaks and where the emission from star
forming galaxies shows a minimum, it can be considered a good proxy for AGN
radiative power. For SMGs, the MIR luminosity is contaminated by the
host-galaxy light, therefore it only provides an upper limit to the AGN
power. MIR luminosities are available for the HzRGs from the
literature~\citep{seymour07}. For the IsOQs, they are estimated using the
models in~\citet{polletta08}. For the SMG/AGNs, they are derived by fitting
their SEDs, including 24\,\micron\ data, with galaxy
templates~\citep[see][]{polletta07}.

The stellar NIR luminosities of all selected sources are compared with the
MIR luminosities in Figure~\ref{lh_lmir}.Note that the lack of IsOQs at L(MIR)$<10^{12}$\,\lsun\ is due to the way
these objects were selected~\citep{polletta08}.

The selected AGNs are characterized by a broad range of NIR/MIR
(stellar/AGN) luminosity ratios. Since the NIR luminosity is a proxy for
stellar mass and the MIR luminosity a proxy for AGN power, the broad range
of L(NIR)/L(MIR) (stellar/AGN) luminosity ratios indicate that the three
classes of selected AGNs emit at different Eddington ratios or follow
different M$_{BH}$-M$_{*}$ relations. These two possible interpretations are
explained below and illustrated in Figure~\ref{lh_lmir}. We consider the
local correlation between the bulge mass and BH mass~\citep{marconi03} and a
possible offset given by $\delta$, i.e. Log(M$_{BH}$) =
8.28+0.96$\times$(Log(M$_{*}$)$-$10.9)+Log($\delta$), where M$_{BH}$ is the
BH mass in \msun, and M$_{*}$ the bulge mass in \msun\ derived from L(H).
The BH mass can be derived from the MIR luminosity assuming an Eddington
ratio (L$^{AGN}_{bol}$/L$_{Edd}$=$\lambda$) and a relationship between
L(MIR) and the AGN bolometric luminosity, Log(L$^{AGN}_{bol}$). The latter
can be estimated using the following relation, Log(L$^{AGN}_{bol}$) =
Log(L(5$\mu$m))+0.36$\pm$0.07, derived from the reprocessed thermal emission
produced by the torus~\citep{polletta08}. This relation might underestimate
the AGN bolometric luminosity because it does not include the AGN absorbed
radiation. For comparison, when using the unobscured AGN template
from~\citet{elvis94a}, which includes optical and far-IR components,
Log(L$^{AGN}_{bol}$) = Log(L(5$\mu$m))+1.16.

Assuming the BH and stellar masses estimated as described above, we derived
the expected NIR and MIR luminosities for different values of Eddington
ratio ($\lambda$) and of the offsets from the local M$_{BH}$-M$_{*}$
relation ($\delta$) and report them in Figure~\ref{lh_lmir}. Note that all
expected MIR luminosities would be higher by a factor 6.3 if the
L$^{AGN}_{bol}$--L(5$\mu$m) relation from~\citet{elvis94a} was adopted. Note
that these relations, as well the estimated masses, are characterized by
large uncertainties. However, the uncertainties are smaller than the offsets
observed among the three AGN samples, and all systematic uncertainties would
shift all the values by equal amounts leaving the observed dispersion and
offsets still present among the three AGN classes. Since our interpretation
is based on such offsets rather than on absolute values, our results are not
affected by these uncertainties.

The IsOQs show systematically higher L$^{AGN}_{bol}$/L$_{Edd}$ ratios,
consistent with being close to Eddington-limited, while HzRGs and SMG/AGNs are
characterized by progressively lower Eddington ratios.
Instead of different Eddington ratios, the different L(MIR)/L(H) luminosity
ratios could be explained by different M$_{BH}$--M$_{*}$ relations with
different offsets from the local one. Assuming that all the selected AGNs are
Eddington-limited would imply a relatively small offset
(0.1$<$$\delta$$<$10) from the local M$_{BH}$-M$_{*}$ relation in IsOQs
and larger offsets in HzRGs and SMG/AGNs (0.01$<$$\delta$$<$0.5) and thus
smaller BH masses than expected from the local relationship at parity of
stellar mass. Indeed,~\citet{borys05} find that SMG/AGNs are
characterized by lower BH masses than expected from the local relationship
assuming that they are Eddington-limited. Alternatively, they would lie on
the M$_{BH}$--M$_{*}$ relation if a lower Eddington ratio was assumed. In
summary, the offset in L(H)--L(MIR) space between HzRGs and AGN/SMGs and the
IsOQs can be attributed to a difference in L$^{AGN}_{bol}$/L$_{Edd}$ ratios
or to different offsets from the local M$_{BH}$--M$_{*}$ relation.

In the next section, we derive the median SED of the three samples and model
them with starburst and obscured AGN templates.

\subsection{Obscured AGN infrared SEDs}

\begin{figure*}
   \centering
   \includegraphics[width=18.5cm]{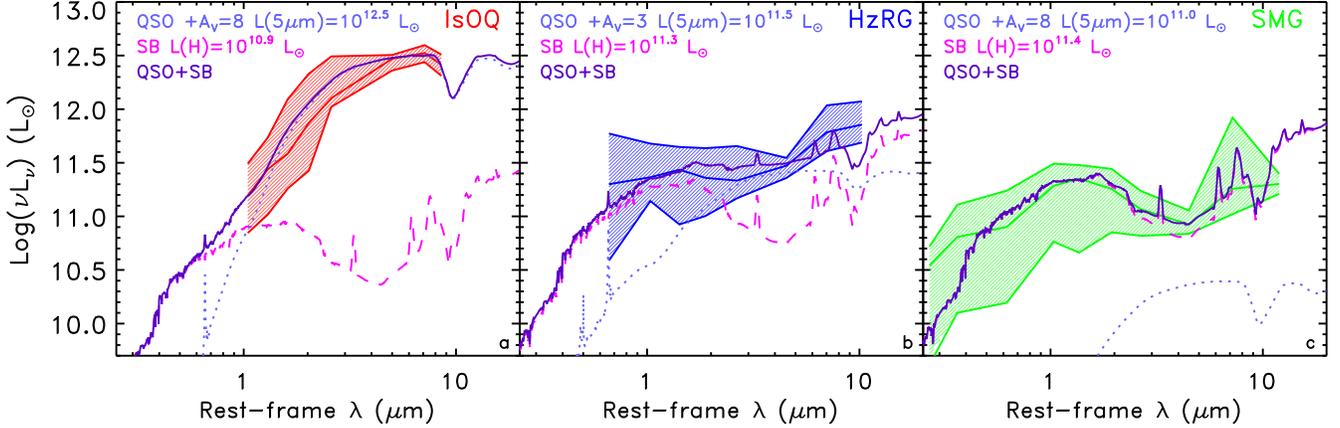}
   \caption{Median SEDs of IsOQs ({\it red line} in panel a), HzRGs ({\it blue
   line} in panel b), and SMG/AGNs ({\it green line} in panel c) normalized
   at their median MIR luminosity.  The shaded areas are obtained from the
   50\% of all points closer to the median SED.  The light-blue dotted line
   represents an extincted QSO template with ${\rm A_{V}}$ as indicated. The
   magenta dashed line represents the starburst (SB) galaxy M\,82 with L(H)
   equal to the median stellar L(H) of each group, as annotated. The purple
   curves represent the sum of the extincted QSO template and of the
   starburst template. The Galactic center extinction curve~\citep{chiar06}
   was applied to redden the QSO template.}
              \label{sed}
    \end{figure*}

In Figure~\ref{sed}, we plot the median SEDs of HzRGs, SMG/AGNs, and IsOQs
after normalizing them at 5\,$\mu$m in the rest frame at the median MIR
luminosity of each group. The median Log(L(5\,$\mu$m)/\lsun) is
12.47$\pm$0.23, 11.55$\pm$0.43, and 10.96$\pm$0.36 for IsOQs, HzRGs, and
SMG/AGNs, respectively. The median and dispersion values are derived by
regrouping the data points of the normalized SEDs in bins with a width of
0.12--0.2 in Log($\lambda$) or larger for less than 4 data points, or
smaller for more than 18 data points. We also show the median template of an
obscured QSO~\citep{polletta07}, and the SED of the prototypical starburst
(SB) galaxy M\,82~\citep{silva98}. The SB template is normalized at the
median stellar H-band luminosity of each group. The median Log(L(H)/\lsun)
is 10.95$\pm$0.32, 11.32$\pm$0.24, and 11.37$\pm$0.21 for IsOQs, HzRGs, and
SMG/AGNs, respectively. The QSO template is added after applying a certain
degree of foreground extinction, \av$<$8, and normalized so that the sum of
the QSO and the SB templates are consistent with the median SED of
each group. We use the Galactic center extinction law by~\citet{chiar06}
extinction curve because it extends to the MIR and has been shown to 
reproduce the IR spectra of AGNs~\citep{sajina07a,polletta08}. Note that the
adopted range of extinction values, \av$\leq$8, corresponds to extreme dust
opacities (up to hundreds of magnitudes) if a proper treatment of absorption
and re-emission is considered as in radiative transfer models.  Thus, the
selected \av\ range represents both the extinction that could be
produced by a dusty galaxy and an optically thick torus. The sum of
the extincted QSO template and of the SB template is also shown in
Figure~\ref{sed}.

Figure~\ref{sed} shows progressively higher stellar luminosities and weaker
AGN radiative power in IsOQs, HzRGs, and SMG/AGNs.
Motivated by recent models that predict feedback on SF and BH growth from AGN-driven radio activity~\citep[e.g.][]{croton06b}, in the
next section, we analyze whether the observed differences in stellar masses
and AGN radiative power are related to the radio properties of the three AGN
classes.

\subsection{Radio properties}\label{radio_prop}

Radio fluxes are available for all HzRGs~\citep{seymour07} and
SMG/AGNs~\citep{borys05}, and for 6
IsOQs~\citep{becker95,polletta06,sajina07b}. The radio luminosities and
MIR/radio luminosity ratios of the 6 IsOQs are listed in Table 1, and their
location shown in Figure~\ref{lh_lmir}.  We also list 
median values and ranges for all 3 classes. For the HzRGs,
we converted the radio luminosities at 3 GHz to 1.4 GHz assuming a power-law
model with spectral index $\alpha$\,=\,$-$0.8~\citep{seymour07}, where
$S_{\nu}\propto \nu^{\alpha}$. It is well-established that 
the radio emission in HzRGs is AGN-powered, while it is mainly
due to SF activity in SMG/AGNs~\citep{chapman04a}. The two classes show very
different radio fluxes, luminosities and L(MIR)/L(Radio) luminosity ratios.
The HzRG L(MIR)/L(Radio) luminosity ratio is $\gtrsim$10, and it is
$\sim$10$^4$ in SMG/AGNs (see Table 1). Based on the ratio between the radio
and the 24$\mu$m luminosity~\citep{donley05}, we find that one source
(LH\_A8) is radio-loud and 2 are radio intermediate (N2\_09, and MIPS22204).
For 3 of the 6 IsOQs with known radio fluxes, SFRs are also available from
the measured far-IR luminosity, LH\_A8, N2\_09, and
MIPS22204~\citep{polletta08}. In all 3 cases, the SFR derived from the radio
luminosity~\citep{condon92,kennicutt98} is a factor of 10 or 100 higher than
derived from the FIR luminosity. This difference indicates that the radio
emission in some IsOQs might be mainly powered by the AGN. However, compared
to HzRGs, IsOQs are less radio luminous, and they show higher MIR/radio
luminosity ratios (see Table~\ref{radio_tab}). The presence of AGN-driven
intermediate radio activity in some IsOQs has also been found by a study on a
sample of MIR selected sources that include some of the IsOQs in our sample
and others at lower MIR luminosity~\citep{sajina07b}. This study claims that
some IsOQs are at the beginning of a radio-loud phase. In conclusion, we
find a variety of radio properties and no significant correlation with the
L(MIR)/L(NIR) ratio in the selected AGN samples.

\subsection{Star formation rates}

Recent evolutionary models also postulate an interplay between AGN activity
and SF~\citep[e.g.][]{silk05}. Interestingly, in spite of the difference in
the stellar content, AGN power, and radio-activity in these systems, all
SMG/AGNs, many HzRGs, and some IsOQs are characterized by high SFRs. The
SFRs of 8 IsOQs, obtained using MIPS FIR (70, and 160\,$\mu$m) data, range
from 600 to 3,000\,\msun\,yr$^{-1}$~\citep{polletta08}. Typical SFRs for
SMG/AGNs range from 200 to 2,000\,\msun\,yr$^{-1}$~\citep{pope06}. HzRGs are
also characterized by high SFRs, up to a few
1,000\,\msun\,yr$^{-1}$~\citep{reuland04,archibald01}. The similarity in
SFRs can be interpreted as lack of evidence for radio activity or for AGN
radiative power as regulators of SFR. This suggests either that AGN feedback
does not take place as radio activity or AGN radiation, that its effects on
the SFR are not observable yet, or that the feedback process does not act
directly on the SFR. A similar result has recently been found by studying a
sample of X-ray selected AGNs whose UV--NIR emission is dominated by stellar
light~\citep{alonso08}. In this study, no evidence of either suppressed or
elevated SF is found in galaxies with AGN activity compared to galaxies of
similar stellar masses and redshifts, suggesting that AGN activity does not
affect the SFR of their host-galaxies.

\section{Discussion and conclusions}

We analyzed the stellar/AGN luminosity ratio, the level of radio activity,
and the SFR in three classes of AGNs at $z$$\sim$2, IsOQs, HzRGs, and
SMG/AGNs, to investigate the existence of a link between AGN-driven radio
activity, the buildup of stellar mass, and BH growth. We find that these
AGNs are characterized by a broad range of stellar/AGN (L(NIR)/L(MIR))
luminosity ratios.  On average, SMG/AGNs and HzRGs have more massive hosts
than the most luminous, obscured QSOs at similar redshifts. Such a broad
range can be explained either by different Eddington ratios or by different
M$_{BH}$-M$_{*}$ relations.

Here, we have considered the possibility of an evolutionary link among these
various AGNs, but we do not rule out the possibility that the various types
of AGNs are the result of different evolutionary histories or physical
states. Assuming that the stellar and BH mass can only grow, any link among
these three classes of AGNs implies either that IsOQs will become less
AGN-luminous and their hosts more massive, as observed in SMG/AGNs, or that
SMG/AGNs will resemble IsOQs, as their AGN becomes more powerful~\cite[see
e.g.][]{borys05,alexander05b}. This last hypothesis is not supported by the
relatively low stellar masses measured in IsOQs. However, it is possible
that our IsOQ sample is biased towards systems with particularly low-mass
hosts, as they are selected on the basis of red optical-IR colors, we cannot
rule out the latter scenario. On the other hand, the former scenario,
evolution from IsOQs to SMG/AGNs, is quite plausible and supported by other
observations. This scenario would imply that the bulk of stellar mass is
still being assembled in IsOQs, while the bulge is at a more advanced stage
in HzRGs and SMG/AGNs. HzRGs and SMG/AGNs might thus represent a later phase
in the AGN evolution than are IsOQs. According to this scenario, fully grown
BHs would already be in place in high-$z$ galaxies before the bulk of
stellar mass was assembled. This scenario is also consistent with
observations of unobscured QSOs at very high-$z$ ($z>$5) in which a fully
grown BH and large reservoirs of molecular gas are present, but where
spheroidal stars have not yet formed~\citep{walter04}. Another supportive
piece of evidence for this scenario is given by the younger stellar
populations found in unobscured AGNs compared to radio
galaxies~\citep{kauffmann07}. However, this result has been obtained by
studying low-$z$ samples where we can expect different processes to dominate
and evolutionary sequences to take place. Better defined samples and more
accurate stellar and BH masses are needed to distinguish the two possible
scenarios or rule out any evolutionary link among the three selected AGN
samples.

Finally, we want to underline the risks of assuming Eddington-limited AGN
emission or the local M$_{BH}$-M$_{*}$ relation in the study of AGN and
galaxy evolution.

\begin{table}
\begin{minipage}[t]{\columnwidth}
\caption{Radio properties of selected AGNs}
\label{radio_tab}
\centering       
\renewcommand{\footnoterule}{}
\begin{tabular}{l c c c}        
\hline\hline                 
Source name    &
   $z$         &
 L$_{1.4\,GHz}$\footnote{\small The radio luminosities are from the deep VLA
survey of the Lockman Hole~\citep[][Owen et al., in prep.]{polletta06} for
LH\_A4, LH\_A5, LH\_A6, and LH\_A8, from the FIRST survey for N2\_09, and
from the VLA-FLS survey~\citep{sajina07b} for MIPS22204. The radio
luminosities for the HzRGs and SMG/AGNs are from~\citet{seymour07}
and~\citet{pope06}, respectively.} &
 Log$\left(\frac{L(5\mu m)}{\nu L_{1.4\,GHz}}\right)$ \\
                 &               & (W Hz$^{-1}$)   &                            \\
\hline\hline
 IsOQ/LH\_A4     &  2.54         & 25.04$\pm$0.02  &  5.27 \\
 IsOQ/LH\_A5     &  1.94         & 24.54$\pm$0.02  &  4.97 \\
 IsOQ/LH\_A6     &  2.20         & 24.42$\pm$0.03  &  5.09 \\
 IsOQ/LH\_A8     &  2.42         & 26.71$\pm$0.02  &  3.22 \\
 IsOQ/N2\_09     &  1.98         & 25.69$\pm$0.03  &  4.15 \\
 MIPS22204       &  2.08         & 25.68$\pm$0.02  &  4.30 \\
 Median(IsOQ)\footnote{\small Includes only the 6 IsOQs with available radio data.
The median redshift for the whole IsOQ sample is 2.23$\pm$0.31, and the
redshift range is 1.29--2.96.}
                 & 2.14$\pm$0.19 & 25.68$\pm$0.68  &  4.97$\pm$0.61 \\
 Range(IsOQ)$^b$ & 1.94--2.54    & 24.42--26.71    &  3.22--5.27 \\
 Median(HzRG)    & 1.90$\pm$0.53 & 27.89$\pm$0.48  &  1.11$\pm$0.37 \\
 Range(HzRG)     & 1.08--3.09    & 26.55--28.53    &  0.60--2.30 \\
 Median(SMG/AGN) & 2.20$\pm$0.43 & 24.31$\pm$0.28  &  4.25$\pm$0.39 \\
 Range(SMG/AGN)  & 0.98--2.58    & 23.25--24.63    &  3.71--5.18    \\
\hline
\end{tabular}
\end{minipage}
\end{table}

\begin{acknowledgements}
We thank the referee for comments that improved the paper. We are grateful to
C. De Breuck for providing HzRG data in electronic format and for comments,
F. Owen for providing some of the radio data for the IsOQs and for comments,
and S. Kassin for helpful discussions. MP acknowledges financial support
from the Marie-Curie Fellowship grant MEIF-CT-2007-042111.

\end{acknowledgements}

\end{document}